%% file: Open_prob_CombPhys_v5.tex
\documentclass[12pt]{iopart}
\usepackage[dvips]{epsfig}
\usepackage{amssymb}
\usepackage{amsfonts}
\usepackage{amscd}
\usepackage{pstricks}
 \font\tensym=msbm10
 \font\sevensym=msbm7
 \font\fivesym=msbm5

 \font\tengoth=eufb10
 \font\sevengoth=eufb7
 \font\fivegoth=eufb5

\newfam\symfam
\textfont\symfam=\tensym \scriptfont\symfam=\sevensym
\scriptscriptfont\symfam=\fivesym

\newfam\gothfam
\textfont\gothfam=\tengoth \scriptfont\gothfam=\sevengoth
\scriptscriptfont\gothfam=\fivegoth

\def\hs{\hbox to 3mm{}}
\def\hhs{\hbox to 5cm{}}
\def\ss{\smallskip}
\def\ms{\smallskip}
\def\bs{\bigskip}

\def\JPicScale{0.8}\ifx\JPicScale\undefined\def\JPicScale{1}\fi

\def\C{\mathbb{C}}

\def\N{\mathbb{N}}

\def\HWS{{\cal H}_{WS}}
\def\1H{\mathbf{1}_{\HWS}}
\def\L{\mathbb{L}}
\def\V{\mathbb{V}}

\def\Q{\mathbb{Q}}

\def\al{\alpha}
\def\be{\beta}

\def\card{\mathrm{card}}
\def\2m#1#2{(#2 #1)}
\def\3m#1#2#3{(#3 #2 #1)}

\def\diag{\mathbf{diag}}

\def\ra{\rightarrow}

\def\adots{\mathinner{\mkern2mu\raise1pt\hbox{.}
\mkern3mu\raise4pt\hbox{.}\mkern1mu\raise7pt\hbox{.}}}

\def\pointir{\unskip . --- \ignorespaces}

\def\up#1{\raise 1ex\hbox{\footnotesize#1}}

\def\mref#1{{\footnotesize ({\ref{#1}})}}

\newtheorem{expl}{Example}[section]

\newtheorem{proposition}[expl]{Proposition}

\newtheorem{resultat}[expl]{Result}

\begingroup
\def\2#1{\ifnum#1<10 0\fi\the#1}
\xdef\isodayandtime{
{\2\day-\2\month-\the\year\space\2{\count0}:%
\2{\count2}}}
\endgroup

\begin{document}

\title{Some Open Problems in Combinatorial Physics}
\author{G H E Duchamp$^{a}$, H Cheballah$^{a}$ and the CIP team.}

\address{$^a$ LIPN - UMR 7030\linebreak
CNRS - Universit\'e Paris 13\linebreak F-93430 Villetaneuse,
France\vspace{2mm}}

\eads{\linebreak\mailto{ghed@lipn-univ.paris13.fr},
\linebreak\mailto{hayat.cheballah@lipn-univ.paris13.fr}}

\isodayandtime

\tableofcontents

\newpage

\section{Problem A: Multiplicities in $\diag$.}
\subsection{Setting}

Let $\mathcal{H}(F,G)$ be the Hadamard exponential product as
defined below by

\begin{eqnarray}
F(z)=\sum_{n\geq 0} a_n\frac{z^n}{n!},\ G(z)=\sum_{n\geq 0}
b_n\frac{z^n}{n!},\ \mathcal{H}(F,G):=\sum_{n\geq 0}
a_nb_n\frac{z^n}{n!}\ .
\end{eqnarray}

In the case of free exponentials, that is if we write the
functions as
\begin{eqnarray}
F(z)=\exp\left(\sum_{n=1}^\infty L_n\frac{z^n}{n!}\right),\ \ \ \
\ \ \ G(z)=\exp\left(\sum_{n=1}^\infty V_n\frac{z^n}{n!}\right)\ ,
\end{eqnarray}
and using the expansion with Bell polynomials in the sets of
variables $\L=\{L_n\}$, $\V=\{V_m\}$ (see \cite{GOF4,OPG} for
details), we obtain

\begin{eqnarray}\label{prod_thru_bell}
\mathcal{H}(F,G)=\sum_{n\geq 0} \frac{z^n}{n!} \sum_{P_1,P_2\in
UP_n} \L^{Type(P_1)}\V^{Type(P_2)}
\end{eqnarray}
where $UP_n$ is the set of unordered partitions of $[1\cdots n]$.

\ss An unordered partition $P$ of a set $X$ is a subset of
$P\subset \mathfrak{P}(X)-\{\emptyset\}$\footnote{The set of
subsets of $X$ is denoted by $\mathfrak{P}(X)$ (this notation
\cite{B_ST} is that of the former German school).} (that is an
unordered collection of blocks, i. e. non-empty subsets of $X$)
such that
\begin{itemize}
    \item the union $\bigcup_{Y\in P}Y=X$ ($P$ is a covering)
    \item $P$ consists of disjoint subsets, i. e.\\
    $Y_1,Y_2\in P\ and\ Y_1\cap Y_2\neq \emptyset \Longrightarrow Y_1=Y_2$.
\end{itemize}

The type of $P\in UP_n$ (denoted above by $Type(P)$) is the
multi-index $(\al_i)_{i\in \N^+}$ such that $\al_k$ is the number
of $k$-blocks, that is the number of members of $P$ with
cardinality $k$.

\ss Let $P_1,P_2$ be two unordered partitions of the same set. To each labelling of the blocks
\begin{equation}
P_r=\{B_i^{(r)}\}_{1\leq i\leq n_r}\ ;\ r=1,2
\end{equation}

one can associate the intersection matrix
\begin{equation}
M=\left(\card(B_i^{(1)}\cap B_j^{(2)})\right)_{1\leq i\leq n_1\ ;\
1\leq j\leq n_2}\ .
\end{equation}
As $(P_1,P_2)$ are, in essence, unlabelled, the arrow so
constructed

\begin{equation}\label{corr1}
    (P_1,P_2)\mapsto class(M)=d
\end{equation}
aims at classes of packed matrices \cite{DHT} under permutations of rows and columns.\\
These classes have been shown \cite{GOF7,GOF8} to be in one to one
correspondence with Feynman-Bender diagrams \cite{BBM} which are
bicoloured graphs with $p$ ($=\card(P_1)$) black spots, $q$
($=\card(P_2)$) white spots, no isolated vertex and
integer multiplicities. We denote the set of such diagrams by $\diag$ \cite{GOF12,FPSAC07}.\\
Then, the correspondence goes as showed below.

\bs
\input{unlabelled2_1}

\bs\bs Noting $mult(d)$ the cardinality of each fibre of
\mref{corr1}, formula \mref{prod_thru_bell} reads

\begin{eqnarray}
\mathcal{H}(F,G)=\sum_{n\geq 0} \frac{z^n}{n!} \sum_{d\in
diag\atop |d|=n} mult(d)\L^{\al(d)}\V^{\be(d)}
\end{eqnarray}

where $\al(d)$ (resp. $\be(d)$) is the ``white spots type'' (resp.
the ``black spots type'') i.e. the multi-index $(\al_i)_{i\in
\N^+}$ (resp. $(\be_i)_{i\in \N^+}$) such that $\al_i$ (resp.
$\be_i$) is the number of white spots (resp. black spots) of
degree $i$ ($i$ lines connected to the spot) and $mult(d)$ is the
number of pairs of unordered partitions of $[1\cdots |d|]$ (here
$|d|=|\al(d)|=|\be(d)|$ is the number of lines of $d$) with
associated diagram $d$.

\subsection{Problem A}
Give a formula (as smart as possible) for $mult(d)$ as a function
of $d$ (in the language of \cite{DHT}, as a function of the class
of a packed matrix under the permutation of rows and columns).

\ss
{\bf Hint}\pointir For practical computations, one of the two partitions 
may be kept fixed, say $P_1$ and the result of the enumeration multiplied 
by $\frac{n!}{|stab(P_1)|}$.

\section{Problem B: Combinatorics of Riordan-Sheffer one-parameter groups.}

We start with the (vector) space $\C^{\N\times \N}$ of complex bi-infinite matrices.\\
Let $\mathcal{RF}(\N,\C)=(\C^{(\N)})^{\N}$ the space of row-finite
matrices (i. e. matrices for which every row is finitely
supported). To every matrix $T\in \mathcal{RF}(\N,\C)$, one can
associate the sequence transformation
\begin{equation}
    (a_k)_{k\in \N}\mapsto (b_n)_{n\in \N}
\end{equation}
given by
\begin{equation}
    b_n=\sum_{k\in\N}T[n,k]a_k
\end{equation}
this sum is finitely supported as $T\in \mathcal{RF}(\N,\C)$. One
can prove that the set $\mathcal{RF}(\N,\C)$
is exactly the algebra of continuous endomorphisms of $\C^{\N}$ endowed with the topology of pointwise convergence.\\
This transformation can be transported on EGFs by
\begin{equation}
    f=\sum_{k\in\N}a_k \frac{z^k}{k!}\mapsto \hat{f}=\sum_{n\in\N}b_n \frac{z^n}{n!}
\end{equation}
and, in case $\hat{f}$ is given by
\begin{equation}\label{subs1}
\hat{f}(z)=\Phi_{g,\phi}[f](z)=g(z)f(\phi(z)).
\end{equation}
with
\begin{equation}
g(z)=1+{\rm higher\ terms}\ and\ \phi(z)=z+{\rm higher\ terms}.
\end{equation}
we say that the matrix is a matrix of substitutions with prefunction.\\
In classical combinatorics (for OGF and EGF), the matrices
$M_{g,\phi}(n,k)$ are known as {\it Riordan matrices} (see
\cite{Roman,Shapiro} for example). One can prove, using a
Zariski-like argument, the following
proposition \cite{OPG,CDP}.

\begin{proposition}\cite{OPG} Let $M$ be the matrix of a substitution with prefunction; then so is $M^t$ for all
$t\in \mathbf{C}$.
\end{proposition}

\subsection{Problem B} a) Provide a combinatorial proof of the preceding proposition for $t\in \Q$
(without using the "pro-algebraic" structure of the group of substitutions with prefunctions, directly or indirectly).\\
b) Give a combinatorial interpretation of $M^{1/2}$ for some
Sheffer matrices.

\section{Problem C: A corpus for combinatorial vector fields.}

With the preceding notations one can show that, if $M$ is a matrix
of substitution with prefunction, the limit
\begin{equation}
    \lim_{q\ra +\infty}q(M^{\frac{1}{q}}-I)
\end{equation}
exists (call it $L$) and the associated transformation of
sequences (see above) is the sum of a vector field and a scalar
field. One can see that
\begin{equation}
    M\in \Q^{\N\times \N}\Longrightarrow L\in \Q^{\N\times \N}\ .
\end{equation}
in addition, if $M$ is a matrix of substitution (i. e. the prefunction is $\equiv 1$) then the scalar field is zero and so the associated differential operator is a pure vector field (with coefficients in $\Q$ if $M$ is in $\Q^{\N\times \N}$).\\
On the other hand, if $\mathcal{C}$ is a class of labelled graphs
for which the exponential formula applies, the matrix $M$ such
that
\begin{equation}
\hspace{-2cm}   M[n,k]=\textit{Number of graphs labelled by
$[1..n]$ and with $k$ connected components}
\end{equation}
is a matrix of substitution \cite{OPG}. For example with the
graphs of equivalence relations on finite sets, the substitution
is $z\mapsto e^z-1$; for graphs of idempotent endofunctions, the
substitution is $z\mapsto ze^z$.

\subsection{Problem C} a) What is the combinatorial interpretation of the coefficients of the vector field for the two preceding examples ?\\
b) Can we give any insight of the form of this vector field for
general classes of graphs ?

\ss
{\bf Hint}\pointir $M^z=e^{zlog(M)}$ where $log(M)$ is the matrix of a differential operator of the form $q(z)\frac{d}{dz}+v(z)$.

\section{Problem D Probabilistic study of approximate substitutions}

Our motivation, in this section, consists in approximating the
matrices of infinite substitutions by finite matrices of
(approximate) substitutions.  We are then interested in the
probabilistic study of these matrices. To this end, we randomly
generate unipotent (unitriangular) matrices and observe the
number of occurrences of
matrices of substitutions.\\

We start by giving some examples of our experiment which are
summarized in the table below:

\begin{center}
\begin{tabular}{|c||c||c||c|}
  \hline
  Size & Number of drawings & Range of variables & Probability \\
  \hline
  \hline
  \hline
   $[3\times 3]$& $300$  & $[1\cdots 10]$ & $1$  \\
   \cline{3-4}
     &  & $[1\cdots 100]$ & $1$ \\
   \cline{3-4}
   &  &$[1\cdots 10000]$ & $1$ \\   \hline
   \hline
   \hline
   $[4\times 4]$& $275$ &  $[1\cdots 10]$ & $0.0473$  \\
   \cline{3-4}
     &  & $[1\cdots 100]$ & $0.0001$ \\
   \cline{3-4}
   &  &$[1\cdots 10000]$ & $0^+$ \\ \hline
   \hline
   \hline
   $[10\times 10]$& $1500$ &  $[1\cdots 10]$ & $0.0327$ \\
   \cline{3-4}
     &  & $[1\cdots 100]$ & $0^+$ \\
   \cline{3-4}
   &  &$[1\cdots 10000]$ & $0^+$ \\
   \hline

\end{tabular}\end{center}

\vspace{0.5cm}

According to the results obtained, we observe that the (approximate) 
substitution matrices are not very frequent. However, in meeting
certain conditions such as size, the number of drawings and the
range of the variables, we can obtain positive probabilities that
these matrices appear.\\ Let us note that the smaller the size of
the matrix the more probable one obtains a matrix of
substitution in a reasonable number of drawings.\\
 We also notice that, if
we vary the range of variables, and this in an increasing way and
by keeping unchanged the number of drawings and size, the
probability tends to zero. We also notice that the unipotent
matrices of size 3 are all matrices of approximate substitutions.
This is easy to see because the exponential generating series of
the 3$^{rd}$ column
 will always have the form $\displaystyle{c_k=\frac{x^2}{2!}}$.\\ Thus, we can say
that the test actually starts   from the matrices of  size higher
or equal to 4.

\begin{resultat}
Let $r$ represent the cardinality of the range of variables and
$n\times n$ be the size of the matrix.\\
According to the results obtained; we can say that the 
probability $p_n$ of appearance of the matrices of substitutions 
depends on $r$ and $n$ and we have the following upper bound:

\begin{eqnarray}
  p_{n} &\leq& \frac{r^{2n-3}}{r^{\frac{n(n-1)}{2}}}
\end{eqnarray}
which shows that

\begin{center}
\begin{eqnarray}
  p_n\longrightarrow 0\quad  &\mbox{as}& \quad n\longrightarrow\infty
\end{eqnarray}
\end{center}

\end{resultat}

\subsection{Problem D} One can conjecture that the effect of the
range selection vanishes when $n$ tends to infinity. More precisely:
\begin{eqnarray}
  p_{n} &\sim& \frac{r^{2n-3}}{r^{\frac{n(n-1)}{2}}}
\end{eqnarray}

\end{document}

%% file: unlabelled2_1.tex
\ifx\JPicScale\undefined\def\JPicScale{1}\fi
\unitlength \JPicScale mm

\begin{picture}(122.5,87.5)(0,5)
\linethickness{0.75mm}
\multiput(69.34,49.61)(0.12,0.37){76}{\line(0,1){0.37}}
\linethickness{0.75mm}
\multiput(72.23,52.11)(0.12,0.37){67}{\line(0,1){0.37}}
\linethickness{0.75mm}
\multiput(82.76,78.82)(0.12,-0.57){48}{\line(0,-1){0.57}}

\put(60,80){\circle{5}}
\put(80.13,79.87){\circle{5}}
\put(100.13,79.87){\circle{5}}
\put(120,80){\circle{5}}

\put(70.2,49.88){\circle*{5}}
\put(90.3,49.56){\circle*{5}}
\put(110,50){\circle*{5}}

\put(56,86){$\{1\}$}
\put(70,86){$\{2,3,4\}$}
\put(88,86){$\{5,6,7,8,9\}$}
\put(115,86){$\{10,11\}$}

\put(60.2,41.88){$\{2,3,5\}$}
\put(77.3,41.88){$\{1,4,6,7,8\}$}
\put(104,41.88){$\{9,10,11\}$}

\linethickness{0.75mm}
\multiput(71.97,49.74)(0.12,0.14){212}{\line(0,1){0.14}}

\linethickness{0.75mm}
\multiput(102.76,79.47)(0.12,-0.48){57}{\line(0,-1){0.48}}

\linethickness{0.75mm}
\multiput(92.76,49.61)(0.12,0.36){81}{\line(0,1){0.36}}
\linethickness{0.75mm}
\multiput(90.26,48.29)(0.12,0.37){78}{\line(0,1){0.37}}
\linethickness{0.75mm}
\multiput(88.29,49.08)(0.12,0.36){81}{\line(0,1){0.36}}
\linethickness{0.75mm}
\multiput(111.97,50.66)(0.12,0.37){72}{\line(0,1){0.37}}
\linethickness{0.75mm}
\multiput(109.74,51.05)(0.12,0.37){73}{\line(0,1){0.37}}
\linethickness{0.75mm}
\multiput(61.71,78.03)(0.12,-0.13){216}{\line(0,-1){0.13}}
\end{picture}

\vspace{-2cm}

{\small{\bf Fig 1}\pointir \it Diagram from $P_1,\ P_2$ (set partitions of $[1\cdots 11]$).\\ 
$P_1=\left\{\{2,3,5\},\{1,4,6,7,8\},\{9,10,11\}\right\}$ and  $P_2=\left\{\{1\},\{2,3,4\},\{5,6,7,8,9\},\{10,11\}\right\}$ (respectively black spots for $P_1$ and white spots for $P_2$).\\ 
The incidence matrix corresponding to the diagram (as drawn) or these partitions is 
${\pmatrix{0 & 2 & 1 & 0\cr 1 & 1 & 3 & 0\cr 0 & 0 & 1 & 2}}$. But, due to the fact that the defining partitions are unordered, one can permute the spots (black and white, between themselves) and, so, the lines and columns of this matrix can be permuted. The diagram could be represented by the matrix ${\pmatrix{0 & 0 & 1 & 2\cr 0 &  2 & 1 & 0\cr 1 & 0 & 3 & 1}}$ as well.}


%% file: Open_prob_CombPhys_v5.bbl
\newpage
\section*{References}
\begin{thebibliography}{ABC}
%
\bibitem{BBM} {\sc C. M. Bender, D. C. Brody, and B. K. Meister},
Quantum field theory of partitions, J. Math. Phys. Vol 40 (1999)
\bibitem{GOF7} {\sc P. Blasiak, A. Horzela, K. A. Penson, G. H. E. Duchamp,
A.I. Solomon}, {\it Boson normal ordering via substitutions and
Sheffer-Type Polynomials}, Phys. Lett. A {\bf 338} (2005) 108 
\bibitem{GOF8} {\sc P. Blasiak,  K. A. Penson, A.I. Solomon,
A. Horzela, G. H. E. Duchamp}, {\it Some useful formula for
bosonic operators}, Jour. Math. Phys.
 {\bf 46} 052110 (2005).
\bibitem{B_ST} {\sc Bourbaki N.}, {\it Theory of sets}, Springer
\bibitem{CDP} {\sc H. Cheballah, G. H. E. Duchamp, K. A. Penson}, 
Approximate substitutions and the normal ordering
problem, Symmetry and Structural Properties of Condensed Matter, 
IOP Publishing Journal of Physics: Conference Series, \textbf{104} (2008).\\
{\tt arXiv: quan-ph/0802.1162}
\bibitem{GOF4} {\sc G. H. E. Duchamp, P. Blasiak, A. Horzela, K. A. Penson, A. I. Solomon}, Feynman graphs and related Hopf algebras, Journal of Physics: Conference Series, SSPCM'05, Myczkowce, Poland.
arXiv : cs.SC/0510041 
\bibitem{DHT} {\sc G. Duchamp, F. Hivert, J. Y. Thibon}, {\it Non commutative
functions VI: Free quasi-symmetric functions and related
algebras}, International Journal of Algebra and Computation Vol
12, No 5 (2002). 
\bibitem{GOF12} \textsc{G. H. E. Duchamp, K. A. Penson, P. Blasiak, A. Horzela, A. I Solomon}, \textit{A Three Parameter Hopf Deformation of the Algebra of Feynman-like Diagrams}
\texttt{arXiv:0704.2522} (to be published). 
\bibitem{FPSAC07} {\sc G. H. E. Duchamp, J. -G. Luque, J. -C. Novelli, C. Tollu, F. Toumazet}, {\it Hopf algebras of diagrams}, FPSAC07.
\bibitem{OPG} {\sc G. Duchamp, A.I. Solomon, K.A. Penson, A. Horzela
and P. B lasiak}, {\it One-parameter groups and combinatorial
physics}, Proceedings of the Symposium Third International
Workshop on Contemporary Problems in Mathematical Physics
(COPROMAPH3)
(Porto-Novo, Benin, Nov. 2003),  J. Govaerts, M. N. Hounkonnou and A. Z. Msezane (eds.), p.436 (World Scientific Publishing 2004)\\
arXiv: {\tt quant-ph/04011262} 
\bibitem{Roman} {\sc  S. Roman}, {\it The Umbral Calculus} (New York:
Academic Press) (1984) 
\bibitem{Shapiro} {\sc L. W. Shapiro}, S. Getu, W.J. Woan and L. Woodson,
The Riordan group {\it Discrete Appl. Math.} {\bf 34} 229 (1991)
\end{thebibliography}
